# In-situ Self-optimization of Quantum Dot Emission for Lasers by Machine-Learning Assisted Epitaxy

Chao Shen[1,2,#], Wenkang Zhan[1,2,#], Shujie Pan[1,3], Hongyue Hao[2,4], Ning Zhuo[1,2], Kaiyao Xin[2,5], Hui Cong[2,4], Chi Xu[2,4], Bo Xu[1,2], Tien Khee Ng[6], Siming Chen[1,2], Chunlai Xue[2,4], Fengqi Liu[1,2], Zhanguo Wang[1,2], and Chao Zhao[1,2,*]

[1] Laboratory of Solid State Optoelectronics Information Technology, Institute of Semiconductors, Chinese Academy of Sciences, Beijing 100083, China

[2] College of Materials Science and Opto-Electronic Technology, University of Chinese Academy of Science, Beijing 101804, China

[3] HS Photonics Co., Ltd., Xiangjiang Science & Technology Innovation Base, Changsha, Hunan 413000, China

[4] Key Laboratory of Optoelectronic Materials and Devices, Institute of Semiconductors, Chinese Academy of Sciences, Beijing 100083, China

[5] State Key Laboratory of Superlattices and Microstructures, Institute of Semiconductors, Chinese Academy of Sciences, Beijing 100083, China

[6] Photonics Laboratory, King Abdullah University of Science and Technology (KAUST), Thuwal 23955-6900, Saudi Arabia

*Email: zhaochao@semi.ac.cn

[#]Equally contributing authors

**Abstract**

Traditional methods for optimizing light source emissions rely on a time-consuming trial-and-error approach. While *in-situ* optimization of light source gain media emission during growth is ideal, it has yet to be realized. In this work, we integrate *in-situ* reflection high-energy electron diffraction (RHEED) with machine learning (ML) to correlate the surface reconstruction with the photoluminescence (PL) of InAs/GaAs quantum dots (QDs), which serve as the active region of lasers. A lightweight ResNet-GLAM model is employed for the real-time processing of RHEED data as input, enabling effective identification of optical performance. This approach guides the dynamic optimization of growth parameters, allowing real-time feedback control to adjust the QDs emission for lasers. We successfully optimized InAs QDs on GaAs substrates, with a 3.2-fold increase in PL intensity and a reduction in full width at half maximum (FWHM) from 36.69 meV to 28.17 meV under initially suboptimal growth conditions. Our automated, in-situ self-optimized lasers with 5-layer InAs QDs achieved electrically pumped continuous-wave operation at 1240 nm with a low threshold current of 150 A/cm² at room temperature, an excellent performance comparable to samples grown through traditional manual multi-parameter optimization methods. These results mark a significant step toward intelligent, low-cost, and reproductive light emitters production.

KEYWORDS: Quantum dots, Machine learning, Molecular beam epitaxy, Real-time control, Laser

## Introduction

Self-assembled quantum dots (QDs) exhibit exceptional properties that make them highly valuable as gain media in light-emitting devices, optical amplifiers, and laser diodes.[1,2] To achieve optimal performance, it is crucial to obtain high-density QDs with a uniform size distribution. Precise control over QD density and size minimizes inhomogeneous linewidth broadening, enhancing optical gain and improving device performance, including lower threshold currents, higher quantum efficiency, and increased output power.[3,4] The Stranski-Krastanow (SK) mode in molecular beam epitaxy (MBE) is commonly employed to grow high-quality QDs.[5] However, the results are influenced by multiple variables, such as substrate temperature, III/V ratio, and deposition amount. Traditional methods for optimizing growth conditions tend to be time-consuming, relying heavily on trial and error and the expertise of the MBE grower.[6,7] Analytical models have also been developed to represent the complex physical processes involved in enhancing the optical performance of QDs.[8-11] These methods typically focus on characterizing and adjusting parameters after the material growth rather than adjusting them in real time, which can result in emission shifts and intensity variations.[12,13] *In-situ* optimization of optical performance during growth is ideal for achieving high-quality active regions and improved emission, an area that remains unexplored.

In theory, an *in-situ* photoluminescence (PL) system could be employed to observe the emission of QDs. However, ultrafast PL is necessary for real-time characterization without interrupting growth. Additionally, a significant challenge lies in the quenching of PL in QDs at growth temperatures as high as 500 °C.[14-16] Reflection high-energy electron diffraction (RHEED) is a powerful tool for rapidly acquiring information with minimal impact on the material during growth. RHEED analysis has demonstrated qualitative similarities to *ex-situ* X-ray diffraction analysis.[17]

It has been used to predict the surface morphology of thin films, facilitating smoother film growth.[18] We have reported the growth of QDs with controlled surface density, aided by real-time RHEED data.[19] Researchers have attempted to evaluate the emission wavelength of QDs using RHEED.[20] Real-time RHEED shows promise for analyzing and achieving QDs with superior optical performance.[21-23]

Machine learning (ML) has become increasingly prevalent in analyzing RHEED results due to the large amounts of data involved. ML can automate the identification of various RHEED patterns that emerge during material growth by recognizing patterns and approximating empirical functions in complex systems.[17,24-27,28] It has been used to predict interstitial oxygen concentration and material growth rates.[28-30] However, these approaches are considered *a posteriori* ML-based method, as they rely on data from completed growths and do not provide parameter optimization suggestions during the growth process.[31] Furthermore, the potential of temporal information from RHEED videos remains underutilised.[25,32,33] A practical model should establish a real-time connection between *in-situ* RHEED and material properties, such as optical performance. Using *in-situ* RHEED data as input, real-time output from ML models can establish a mapping relationship between model predictions and material optimization parameters. This enables the optimization of growth parameters during the process, known as *in-situ* control.[21,22] It can quickly identify and correct deviations from expected values, thus improving growth outcomes.[21-23] This significantly enhances optical properties of InAs QDs based on real-time data and model outputs.

In this work, we successfully grew InAs QDs with high PL intensity and narrow full width at half maximum (FWHM) by integrating *in-situ* RHEED with ML. We analyzed the optical characteristics of QDs corresponding to different RHEED patterns and established a real-time correlation between growth status and RHEED data to build a database for an ML model. A lightweight ResNet-GLAM model was designed for the real-time processing of RHEED data, enabling efficient identification of optical performance outcomes. Our results demonstrate that ML can effectively model and predict post-growth optical performance, allowing real-time monitoring and adjusting growth parameters. This model was successfully deployed in our self-developed program, facilitating the growth of single- and 5-layer InAs QDs. The *in-situ* control method proved highly effective in optimizing the optical properties of InAs QDs, resulting in samples with improved PL intensity and narrower FWHM. Ultimately, we achieved high-performance lasers with 5-layer InAs QDs as the active region. The performance of these lasers was comparable to those produced using traditional manual optimization of multiple parameters. Our approach enables *in-situ* characterization and optimization of parameters during material growth, marking a significant advancement in achieving precise control over material growth. This method has the potential for large-scale production, reducing optimization cycles and improving final yield.

Results

InAs/GaAs QD lasers were grown on n-GaAs substrates using a solid-source MBE system. The structure comprises the active region, top and bottom waveguide, cladding, and contact layers, as shown in Figure 1a. The successful growth of QDs in the active region was achieved through dataset construction, model training, and deployment. The size, shape, and density of the QDs, which subsequently impact their optical performance, were determined by the substrate temperature, the InAs deposition amount, and the V/III ratio etc. During our material growth, the indium (In) cell's temperature is maintained constant to achieve a stable growth rate for InAs.[34,35] A series of control samples were grown at an InAs growth rate of around 0.014 ML/s with different V/III ratios (see Supplementary Information for the result of InAs QDs growth with different V/III ratios, S1). The optimal V/III ratio for growing InAs QDs was selected and fixed for subsequent growth. Analysis of the RHEED patterns collected at different InAs deposition amounts and PL results reveals that samples with the best optical performance exhibit faintly discernible chevron streaks, as shown in Figure 1b (see Supplementary Information for the result of InAs QDs growth with different InAs deposition amounts, S2).[36,37] In contrast, a reduction in PL intensity is observed when the chevron streaks are visible. To help the model identify RHEED features associated with the best optical performance, data collected 10 seconds before and after the completion of growth for the sample with the highest PL intensity and the narrowest FWHM were labelled "Yes" to indicate the optimal growth conditions during this period. At the growth rate of 0.014 ML/s, the variation in InAs deposition during this 10-second data collection is less than 0.15 ML; the performance variation of the samples grown under such slight deposition fluctuation is relatively small and can be ignored.[44-46] This approach can effectively prevent issues such as category imbalance, evaluation index imbalance, overfitting, and poor model generalization ability due to the limited amount of data corresponding to the "Yes" label when creating the dataset. All other data from initial growth to 10 seconds before the completion of growth were labelled "No" to signify that they did not meet the optimal growth conditions.

We also found that transitions between distinct reconstructions can be observed *via* RHEED as the QD growth temperature and the deposition amount vary.[38-41] This indicates that RHEED can serve as a predictive tool for determining the material state, correlating with growth temperature and deposition amount.[42,43] Maintaining the substrate within an optimal temperature range ensures uniform QD size, moderate density, and high crystal quality. As shown in Figure 1c, PL testing was performed on samples of the same structure, grown on different quarter wafers of a 2-inch n-GaAs substrate with varying InAs QD growth temperatures. As the substrate temperature gradually decreases, InAs QDs' PL intensity increases from 3049.5 to 3894.4 and then decreases to 2533.2. The FWHM for each spectrum also shows a trend of narrowing from 41.0 to 33.4 meV and then broadening to 41.8 meV. At low temperatures, the reduced mobility of adatoms leads to uneven stacking and excessive density, resulting in the formation of aggregates that induced nonradiative recombination, ultimately degrading the laser's performance. On the other hand, at high temperatures, excessive diffusion or dissolution of QDs occurs, leading to lower density or larger sizes, which weakens the laser's gain. Therefore, it is essential to maintain the substrate temperature in the proper range during the InAs QDs growth to ensure optimal optical performance. So, different labels can be assigned to the data based on the abovementioned characteristics. Among the 4 samples, the one with the best optical performance was selected, and its RHEED data were labeled as "Suitable", indicating the optimal substrate temperature. If the growth temperature was lower than that of the optimal sample, the RHEED data were labelled "Low", indicating insufficient substrate temperature. Conversely, if the growth temperature was higher than that of the optimal sample, the RHEED data were labelled "High", signifying excessive substrate temperature.

We also observed RHEED patterns from the sample changed with the growth temperature of InAs QDs.[38-40] Figure 1d-1i shows typical RHEED patterns captured at different temperatures during the initial phase of QD growth, alongside corresponding AFM images taken after growth completion. At lower temperatures, the samples show a ×1 reconstruction line; at optimal temperatures, a ×2 reconstruction line appears; while at higher temperatures, a ×4 reconstruction line is observed. Additionally, the QD density decreases from $8.6 \times 10^{10}$ cm$^{-2}$ at low growth temperatures to $1.4 \times 10^{10}$ cm$^{-2}$ at high growth temperatures.

The dataset consisted of a total of 24 samples. We applied various image augmentation techniques to enhance the model's ability to generalize. These techniques included adjusting color attributes through random changes in brightness, contrast, saturation, and hue. By expanding the training data with these enhancements, we made the model more resilient to different lighting conditions and color variations, thereby reducing the potential impact of degraded RHEED fluorescent screens. Additionally, we implemented a custom image cropping strategy that involved randomly selecting and adjusting the position and size of a specific image region. This spatial distribution alteration helped the model adapt to various image regions and improved its ability to recognize and interpret image content.

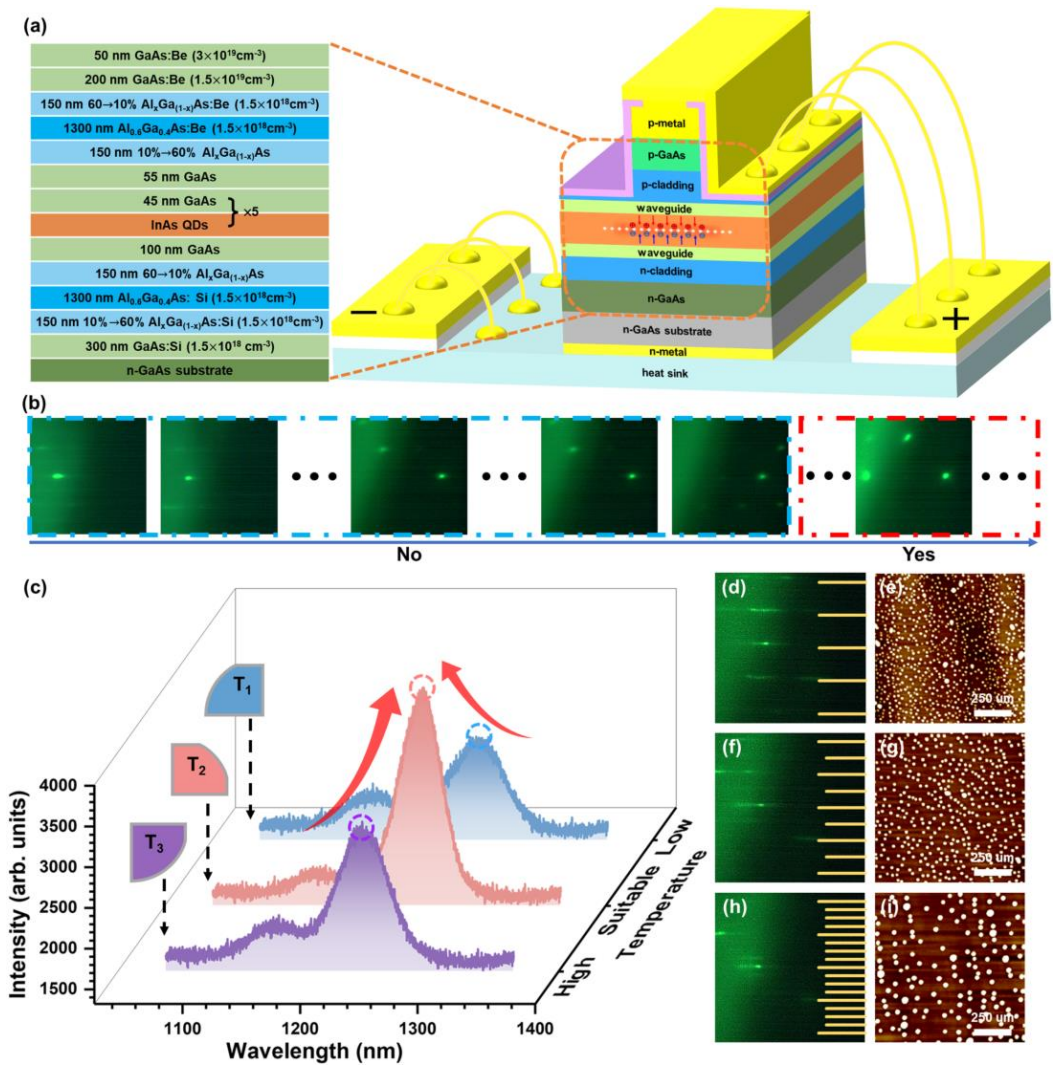

**Figure 1.** (a) The structure of the laser. (b) Evolution of RHEED patterns during the InAs QDs growth process. (c) PL spectra of samples grown at various substrate temperatures. (d-i) Representative RHEED patterns and AFM images of InAs QDs grown under low, suitable, and high-temperature conditions.

The model in this study features a sophisticated architecture specifically designed to extract and utilize both spatial and channel-specific features from processed RHEED data, improving its ability to recognize complex patterns. The architecture consists of three main components: a ResNet block, a Global and Local Attention Modules (GLAM) block, and a Multi-Layer Perceptron (MLP) classifier, as shown in Figure 2a. Two models were trained in this study: the "Temperature Model" and the "Shutter Model". The "Temperature Model" produces three outputs: "Low", "Suitable", and "High", which correspond to the varying growth temperatures of the InAs QDs as discussed earlier. In contrast, the "Shutter Model" has two outputs: "Yes" and "No", indicating whether the sample meet the optimal status identified in the previous analysis. The ResNet block progressively extracts and refines features through residual connections, allowing the model to reduce spatial dimensionality while preserving essential information, as shown in Figure 2b.[47,48] The ResNet block is designed with convolutional layers, residual connections, and normalization operations, allowing the model to learn spatial representations from the input frames adaptively. Each block uses a convolutional layer to extract spatial features, followed by normalization and non-linear activation functions to enhance feature discrimination. The residual connections in the ResNet block facilitate efficient gradient flow, enabling deeper feature extraction without the risk of vanishing gradients.

Before training the model with the entire dataset, a small batch of data was utilized to efficiently optimize various parameters. These parameters included the step size in the convolutional layers of the ResNet Block, the number of repetitions for both the ResNet and GLAM Blocks, and the model's input size, as shown in Figure 2c (see Supplementary Information for the detailed information of modules in the GLAM block, S3).[49-52] The GLAM block is an essential component that combines local and global channel attention mechanisms to refine feature representations. It consists of two parallel branches: global attention, which includes global channel attention (GCA), and global spatial attention (GSA) to capture long-range dependencies and highlight the most significant global features across the input frames. In contrast, local attention includes local channel attention (LCA) and local spatial attention (LSA) to focus on fine-tuning attention at a more localized scale. By merging these two, GLAM significantly improves feature representation by integrating global context and local details, providing a comprehensive focus on the relevant information in the input data. Finally, the model's output is generated by a classical MLP classifier, which processes the aggregated feature maps to carry out the final classification, completing the forward pass of the model, as shown in Figure 2d.

The "Temperature Model" and the "Shutter Model" structures have been optimized, as shown in Figure 2e-2f. The models were pre-trained by modifying the ResNet block to adjust the input data downsampling rate and the number of stacked ResNet and GLAM blocks. The results show that training and validation accuracy improved in the range of 0.25 to 3. This suggests that changing the input size from (N, H, W) to $(2^L N, H/4^L, W/4^L)$ with each ResNet block, along with stacking three ResNet and GLAM blocks, leads to better model performance. This configuration enables the model to focus on both local and global features progressively. Additionally, varying the input pixel size throughout the same training epoch revealed that model performance improves as the input pixel size increases, as shown in Figures 2g-2h. However, the "Shutter Model" does not significantly improve at an input size of 256 pixels, likely due to increased model complexity and insufficient epoch during fast training. As both models are computationally efficient and suitable for hardware implementation, 256-pixel samples were ultimately selected as the input size for both models (see Supplementary Information for training and validation speed of the two models, S4).

The "Shutter Model" focuses on the central area of input data in regions where QDs have not yet formed, as shown in Figure 2i. Once QDs form, the model effectively isolates speckle features and chevron streaks from the original input data, as shown in Figure 2j. Gradient-weighted Class Activation Mapping (GradCAM) analysis further reveals that the "Shutter Model" focuses on subtle features surrounding the specular spot, aligning well with features identified by human observers, demonstrating good model interpretability, as shown in Figure 2k.[53,54] Additionally, random screening of data followed by t-distributed stochastic neighborhood embedding (t-SNE) analysis shows a clear boundary between the "Yes" and "No" labels in a 2D scatter plot, indicating the high sensitivity and accuracy of the "Shutter Model", as shown in Figure 2l.[55-57] Characterizing the "Temperature Model", the feature maps reveal that the model focuses on reconstructing the material surface, specifically streak features. Figures 2m-2o illustrate the ×1, ×2, and ×4 reconstruction under different labels before the QD formation. Additionally, the t-SNE analysis of the "Temperature Model" reveals a clear separation between the scatter points corresponding to different labels, indicating effective discrimination by the model, as shown in Figure 2p. Therefore, when analyzing RHEED data that exhibits periodic changes, the ResNet-GLAM model effectively handle the data even when critical features appear only in individual frames. The ResNet block emphasizes features in individual frames, while the GLAM block focuses on both the sequence of frames and the features of those frames. This approach ensures the model emphasizes the most relevant features, resulting in robust feature extraction and interpretation.

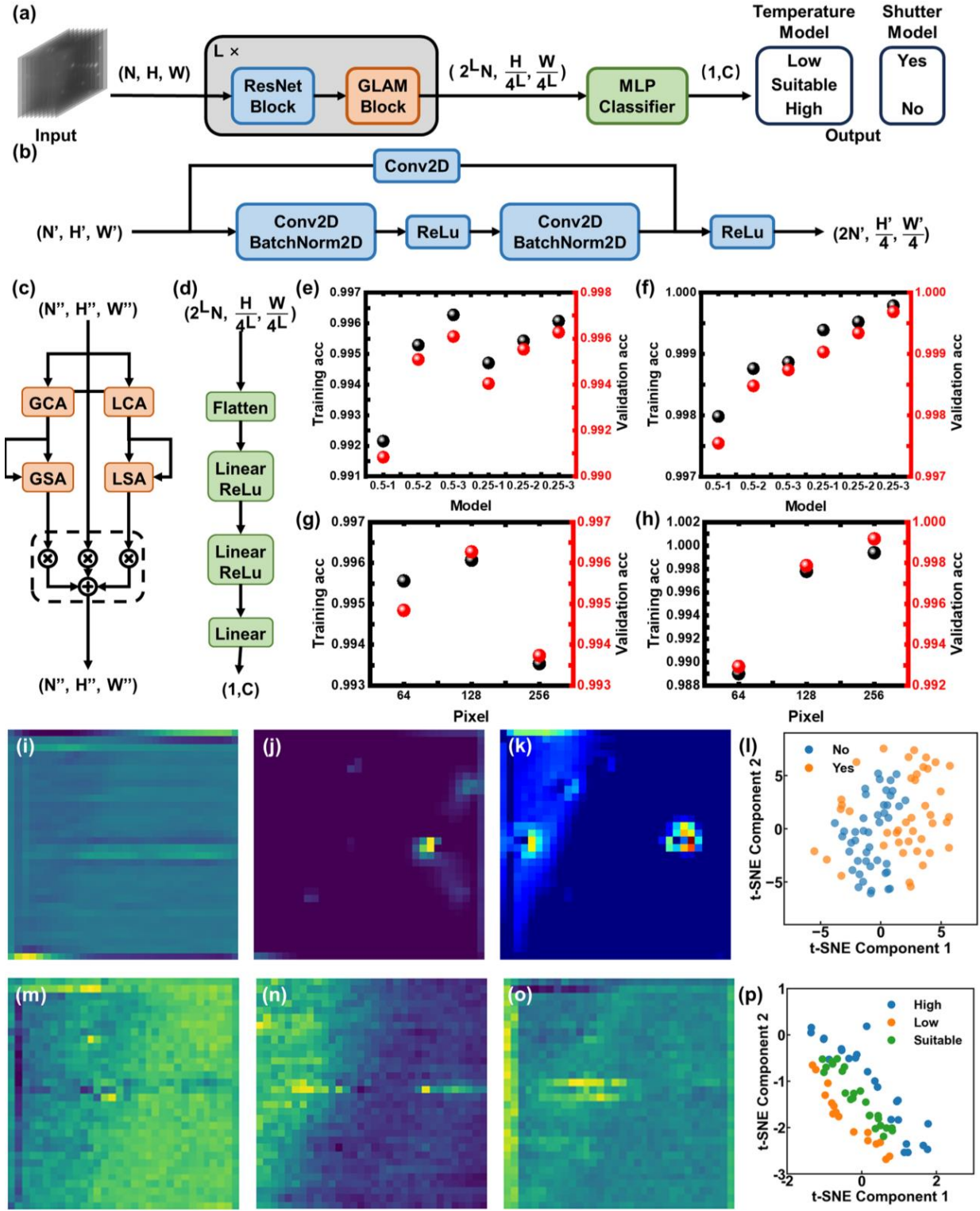

**Figure 2.** (a) Schematic of model inputs and outputs. Architecture of the (b) Adaptive Block, (c) GLAM Block. LCA: local channel attention. LSA: local spatial attention. GCA: global channel attention. GSA: global spatial attention. (d) Structure of the MLP Classifier. (e-h) Training and validation performance of the Shutter and Temperature models with varying model architectures and input data sizes. Characterization results of the (i-l) Temperature model, (m-o) Shutter model.

The ML model, deployed in LabVIEW, optimizes the substrate temperature and InAs deposition amount in real-time by analyzing collected RHEED data, as shown in Figure 3a (see Supplementary Information for the experiment setup and program interface, S5). It is important to note that the substrate rotates continuously, leading to periodic variations in the RHEED videos collected from different angles. During the data processing of the RHEED video, the multi-channel RHEED frames captured in real-time are first converted into arrays of individual luminance channels. These arrays are then combined along additional dimensions to form a 3D array for processing by the ML model, as shown in Figure 3b. Two models, the "Temperature Model" and the "Shutter Model", have been developed in LabVIEW to monitor the QD growth process. These models evaluate and adjust the substrate temperature until the material reaches an optimal state for optical performance, as shown in Figure 3c. Based on the outputs of these models, LabVIEW then adjusts the substrate temperature and the shutter (see Supplementary Information for details on setting the substrate temperature ramp rate, S6). If the "Temperature Model" output indicates "High", the current substrate temperature is above the suitable growth temperature and needs to be decreased. Conversely, if the output is "Low", the substrate temperature must be increased. If the output is "Suitable", the current substrate temperature is close to the ideal level and should remain unchanged to achieve the best optical performance. The control mechanism will continuously adjust the shutter and substrate temperature until the "Shutter Model" outputs "Yes", indicating that the InAs QDs have achieved the optimal state for optical performance, signaling the end of the InAs QD growth process.

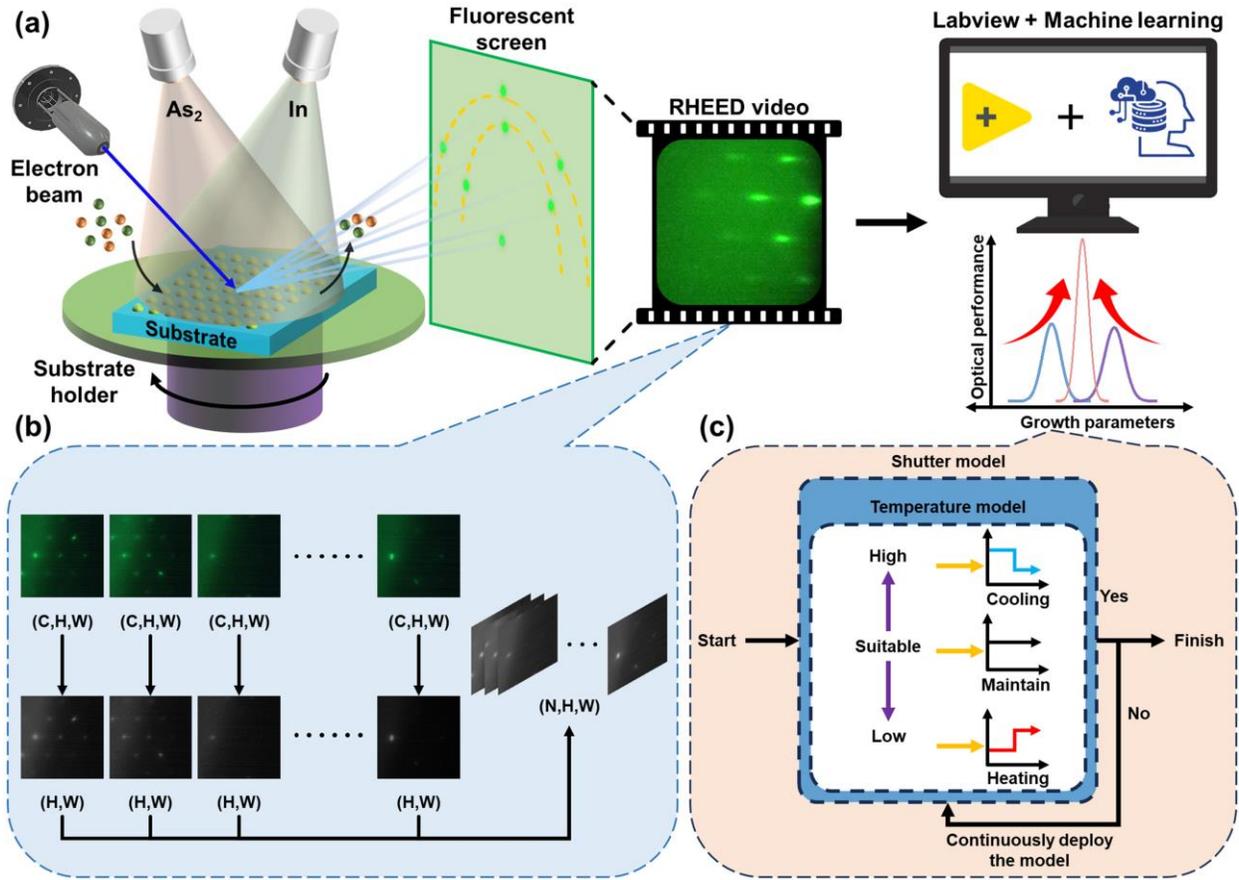

**Figure 3.** Schematic of the (a) experimental data acquisition and material growth system, (b) RHEED data preprocessing workflow, and (c) growth control logic architecture.

After extensive training, we integrated the model into our self-developed LabVIEW program to guide the growth of QDs. To demonstrate the significant improvement in optical quality achieved through the ResNet-GLAM model optimization, we intentionally set the initial growth temperature far from the optimum in each experiment. Before experiment, we grew a control sample A, which exhibited a PL intensity of only 1900.6. This sample highlighted that the growth temperature used was inappropriate for achieving high optical performance (see Supplementary Information for the result of control sample A, S7). Besides, we prepared a control sample B with single-layer InAs QDs, which was manually optimized multiple times. This sample showed significantly improved optical performance, with a PL intensity of 4409.7 and a FWHM of 36.69 meV (see Supplementary Information for the result of control sample B, S8). To demonstrate the effectiveness of our ML model in optimizing growth conditions for improved optical performance, we successfully grew a single-layer buried InAs QDs with PL intensity of 14066.3 and an FWHM of 28.17 meV (see Supplementary Information for the data on single-layer buried QDs, S9). These results represent a 3.2-fold increase and a significant enhancement compared to control sample B.

Figure 4 presents data from a 5-layer buried QD structure for lasers prepared using the ResNet-GLAM model and controlled by the LabVIEW program. The data from each layer of InAs QD growth process is statistically analyzed, encompassing variations in growth temperature and model label output. Figure 4a shows that during the first deployment of the model, the growth temperature changed by 21 °C, indicating that the initial growth temperature was not suitable for producing samples with optimal optical performance. Subsequent model deployments showed no change in substrate temperature, suggesting that the temperature after the first model deployment was appropriate for sample growth. The growth time shows minimal fluctuations, with a variation of only 1.87% for each InAs QD growth, as shown in Figure 4b, indicating a stable InAs deposition amount. In addition, the analysis of the label probability output from the "Temperature model" reveals that, during the initial deployment, the probability of the "High" label was significantly higher than the others, as shown in Figure 4c. However, from the second deployment onward, the probability of the "Suitable" label becomes substantially higher than both the "High" and "Low" labels. This indicates that the model-guided optimization process successfully identifies a suitable growth temperature for developing QDs with excellent optical performance.

In each model deployment, we selected typical RHEED images from the early stages of QD growth during the first, fourth, and final deployments, as shown in Figures 4d-4f. A clear ×4 reconstruction feature was observed during the first model deployment, while a ×2 reconstruction feature was evident in both the fourth and the final model deployments. This confirms that the initial growth temperature of InAs QDs was too high, whereas the temperature during the later deployments was more suitable. In addition, RHEED images were captured during the final stages of growth for each buried InAs QD, as shown in Figures 4g-4i. These images clearly show distinct chevron streak features around the specular spot, confirming that the model accurately identified the optimal time to complete growth.

After growth, the samples were characterized by *ex-situ* AFM and room-temperature PL. The 1 μm × 1 μm AFM image in Figure 4j showed a uniform distribution of QDs, with a density of $4.8 \times 10^{10}$ cm$^{-2}$. The corresponding PL intensity reached 14189.0 with an FWHM of 34.30 meV, superior to control sample C with 5-layer InAs QDs with a PL intensity of 4310.3 and an FWHM of 36.24 meV, which was grown manually based on the parameters of the control sample B (see Supplementary Information for the result of control sample C, S10), indicating minimal dot inhomogeneity and an intensity level comparable to samples optimized through multiple rounds of manual adjustments, as shown in Figure 4k. This shows that the samples prepared using this method exhibit excellent performance.

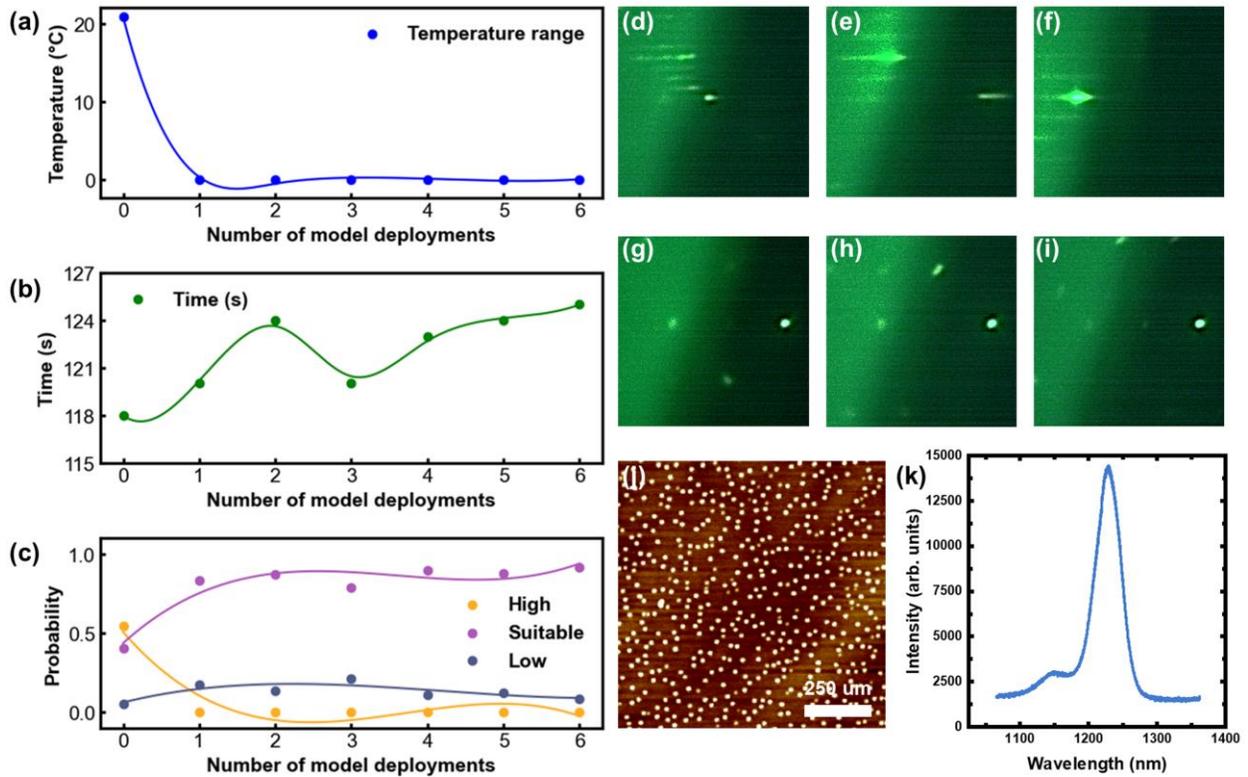

**Figure 4.** The data of 5-layer buried QDs: (a-c) Substrate temperature, growth time, and the output probability of the "Temperature model" during InAs QDs growth. (d-i) Representative RHEED images captured at the initial and near-completion stages of growth during the first, fourth, and final model deployment. (j) The 1 μm × 1 μm AFM image. (k) PL spectrum of the sample.

The method was employed to grow lasers with 5-layer InAs QDs as the active region (see Supplementary Video). Figure 5a shows a high-angle annular dark-field scanning transmission electron microscopy (HAADF-STEM) image of the laser structure, confirming that the layer thicknesses align precisely with the design shown in Figure 1(a). Figures 5b and 5c show the STEM images of the 5-layer InAs/GaAs QD layers and a dark field-TEM of a single QD, respectively (see Supplementary Information for TEM images of the 5-layer InAs QDs, S11). The typical size of InAs/GaAs QD is ~27 nm in diameter and ~8 nm in height. Furthermore, as shown in Figure 5a and Figure 5b, no defects were observed in the active region over a large area. Figures 5d and 5e show the energy dispersive spectroscopy (EDX) elemental mapping for In and Ga, showing a uniform elemental distribution with minimal intermixing. This defect-free and minimal intermixing active region indicates the feasibility of achieving high-quality materials through *in-situ* self-optimization.

The laser structure was processed into broad-area lasers without facet coatings (see Device Fabrication). Figure 5(f) shows typical light-current-voltage (L-I-V) curves of lasers with a 4.36 mm cavity length and 50 µm ridge width, measured in continuous wave operation at 25 °C. The I-V curve exhibits a turn-on voltage of 2.2 V and a differential series resistance of 2.8 $\Omega$ , indicating good metal contacts for efficient current injection. A noticeable kink behavior is observed in the L-I curve at the lasing threshold current density of 150 A/cm², equivalent to 30 A/cm² for each QDs layer. Furthermore, a single facet output power of 16.5 mW was measured at an injection current of 500 mA, resulting in a slope efficiency was 0.095 W/A. These performances are comparable to InAs QDs laser grown on GaAs substrates operating within the same wavelength range.[58,59] The inset of Figure 5(f) shows the SEM image of the fabricated InAs QD laser, which exhibits a mirror-like facet that minimizes the mirror loss due to incomplete cleaving. To reduce the number of longitudinal modes, we selected a laser with a shorter cavity length of 2 mm for spectral testing.[60] Figure 5(g) shows the evolution of emission spectra from a 2 mm cavity length and 50 µm ridge width laser at 25 °C, under various continuous wave operation injection current densities. Below the threshold, the spectrum shows a broad spontaneous emission peak at 1240 nm and a FWHM of approximately 60 nm. Lasing action begins from the spontaneous emission spectrum once the injection current density reaches 290 A/cm². As the current density increases, a distinct multimode laser spectrum emerges, a typical characteristic of broad-area laser structures.

The laser spectrum observed at 1240 nm aligns closely with the wavelength in the PL spectrum, confirming the effectiveness of the *in-situ* self-optimization growth through ML. Figure 5(h) illustrates the collected intensity (L–I) and the linewidth of the lasing peak under various injection current densities, providing evidence of lasing through a noticeable kink in L–I curve and a spectral linewidth narrowing effect. Figure 5(i) shows the device's three-dimensional near-field intensity profile image at an injection current density of 360 A/cm². Two-dimensional and three-dimensional near-field intensity profile images of laser spots were also collected at various current densities (see Supplementary Information for the two-dimensional and three-dimensional near-field intensity profile data, S12). Before reaching the lasing threshold, the emission is predominantly spontaneous, displaying a relatively uniform intensity distribution. Once the lasing threshold is reached, stimulated emission takes precedence, resulting in multiple transverse modes. These results demonstrate that automated and *in-situ* QD laser self-optimization successfully achieves electrically pumped lasing.

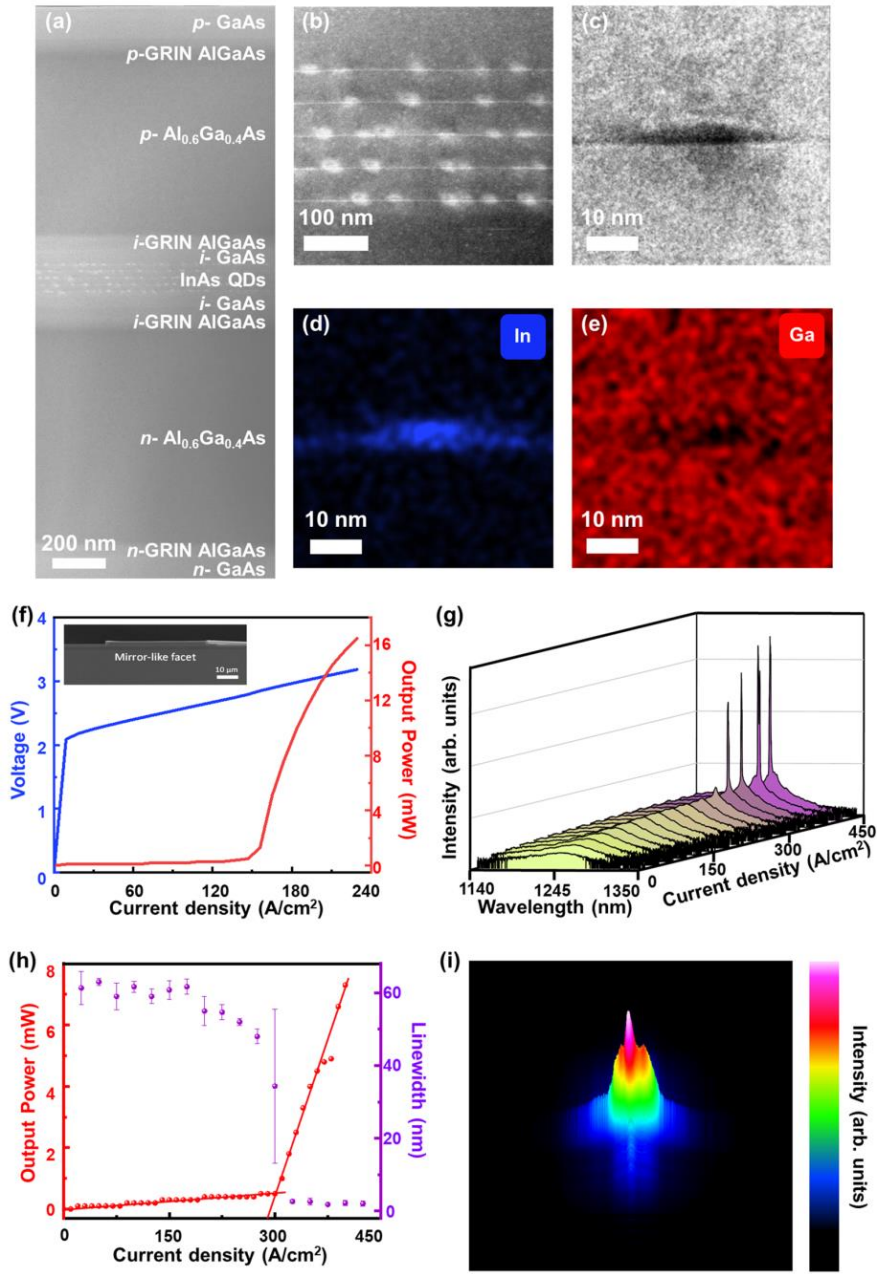

**Figure 5**. Epitaxial growth and structural characterization of InAs/GaAs QD lasers. HAADF-STEM images of (a) the laser, (b) the active layer, and (c) a single QD. (d-e) EDS mapping of In and Ga, respectively. (f) LIV characteristics for a QD laser operated continuous wave at 25 °C. (g) Emission spectra for a QD laser at different injection currents under continuous wave operation at 25 °C. (h) Measured output power (red dots) and linewidth (purple dots) as a function of the injection current density. The error bars represent the standard deviation from multiple measurements. (i) Three-dimensional near-field intensity profile at an injection current density of 360 A/cm² of the QD laser.

## Discussion

The ability to self-optimize laser emission in real-time using ML and feedback control marks a significant advancement in adjusting the optical properties of materials and devices during their preparation. Our findings demonstrate that it is now possible to characterize the optical properties of materials at high temperatures without any obstruction, enabling on-site emission adjustment in emitters based on various material systems. Notably, we have achieved continuous-wave lasing with a low threshold of 150 A cm$^{-2}$ and an output power exceeding 16.5 mW at room temperature, which represents a strong performance for QD lasers in this wavelength range.

Faster response times can be achieved by upgrading GPUs and improving communication protocols between controllers and computers. These advancements facilitate a higher InAs growth rate, leading to a higher density of QDs. Furthermore, the emission wavelength can be extended by equipping the MBE with extra In cells to cap the QDs. It is crucial to emphasize that this work focuses on the self-optimization of the laser's active region. Anticipating even better device performance, we can also optimize other epilayers, adopt industrial-standard fabrication techniques, and apply facet coatings, among other improvements. Our strategy may also enhance other properties, such as carrier mobility, by correlating with the *in-situ* characterization result, thereby improving the high electron mobility transistor (HEMT) performance.

Our demonstration of the capability to grow high-quality III–V materials with tailored characteristics, along with the fabrication of electrically pumped lasers operating in continuous wave mode, opens new avenues for precise control over material growth. With further improvements in hardware, customized modelling, and other areas, this technology holds significant potential for large-scale production, which could enhance productivity and yield in the semiconductor industry.

## Methods

### Material growth

The samples were grown on n-GaAs substrates in a Riber 32P MBE system. The system was equipped with an arsenic (As) valved cracker, indium (In), and gallium (Ga) effusion cells. $As_2$ was used in the growth process, with the As cracker temperature maintained at 900 °C. Beam equivalent pressure (BEP) was used to evaluate the fluxes and calibrate elements III and V ratios.[61] The BEP of In was $5.7 \times 10^{-9}$ Torr, while the BEP of As used to grow GaAs, LT-GaAs, and InAs was $2.5 \times 10^{-6}$, $1.5 \times 10^{-6}$, and $9.6 \times 10^{-6}$ Torr, respectively. C-type thermocouples measured substrate temperatures and the cell's temperatures, and growth rates were calibrated by RHEED oscillations of the additional layers grown on the GaAs substrate. The substrates were outgassed in a buffer chamber at 350 °C and then transformed into the growth chamber. Before the growth of the 300 nm n-doped GaAs contact layer, oxide desorption of the substrates was carried out at 620 °C for 10 minutes. The QD laser structure consisted of the following layers: a 150 nm n-AlGaAs graded cladding layer (graded from 10% to 60% Al), a 1.3 μm lower $Al_{0.6}Ga_{0.4}As$ cladding layer, a 150 nm lower undoped AlGaAs graded-index layer (graded from 60% to 10% Al), an active region containing five layers of InAs QDs, a 100 nm lower and upper GaAs waveguide, a 150 nm upper undoped AlGaAs graded-index layer (graded from 10% to 60% Al), a 1.3 μm upper p-doped $Al_{0.6}Ga_{0.4}As$ cladding layer, another 150 nm AlGaAs graded cladding layer (graded from 10% to 60% Al), and finally, a 250 nm p-doped GaAs contact layer. Si and Be doping levels were varied between $1.5 \times 10^{18}$ cm$^{-3}$ and $3 \times 10^{19}$ cm$^{-3}$ for the bottom and top cladding, respectively. The doping profile remained consistent from the buffer to the undoped waveguide to ensure uniform electrical and optical confinement. The top cladding doping was gradually increased from the top of the undoped waveguide to the contact layer, maintaining the same doping range. This design aimed to prevent Si segregation near the active layer and Be out-diffusion, which can increase free carrier absorption losses. The high doping levels in the cladding improved electron and hole transport, thereby reducing the series resistance in the high Al-content cladding layers. In the active region, InAs QDs were grown at a low growth rate of approximately 0.014 monolayers per second (ML/s) at a low growth temperature. These QDs were capped with a 10 nm low-temperature GaAs (LT-GaAs) layer and a 35 nm GaAs spacer layer to ensure good surface morphology. The LT-GaAs growth was performed using the same As flux and temperature as for the QDs. The GaAs and Al-containing layers in the laser structure were grown at 600 °C and 615 °C, respectively. An uncapped surface InAs QDs reference sample was also grown to characterize the QD morphology.

**Material characterization**

A RHEED electron gun and a fluorescent screen were installed in the MBE growth chamber. The RHEED system (RHEED 12, STAIB Instruments, Inc.) operated at 12 kV and 1.49 A, enabling the epitaxial layer surface analysis during growth. The fluorescent screen was equipped with a shutter, allowing us to capture RHEED video by manually opening it. Additionally, a darkroom with a camera was mounted outside the fluorescent screen and was set with the green light enhancement. The camera's exposure time was set to 100 ms, enabling the capture of RHEED video at a rate of at least 8 frames per second during the substrate rotating at 20 revolutions per minute (RPM). AFM measurements were conducted in tapping mode to enhance surface characterization (Dimension Icon system from Bruker). The custom-built photoluminescence (PL) system comprises an optical beam splitter, reflector, attenuator, and a 532 nm continuous wave excitation laser. The excitation laser is directed to the samples through the optical beam splitter, and an InGaAs detector collects the emission signals in the spectrometer (iHR 550 spectrometers from HORIBA). The setup produced a laser spot approximately 2 mm in diameter on the sample surface, resulting in a laser power density of 15.92 W/cm². During testing, a peak was observed at 1064 nm in the spectral results. This peak can be attributed to the nonlinear optical effects of the second harmonic associated with the 532 nm laser.

The TEM sample was prepared using a Helios 5CX dual-beam FIB microscope from ThermoFisher Scientific. First, a protective layer of Pt was deposited on the exposed target surface inside the FIB microscope. The target area was then FIB ablated at 30 kV to create a prism-shaped structure (10 um length × 3 um width × 5 um depth) along the [011] direction. The resulting sample was mounted onto a Cu TEM grid and milled to create an 80 nm thin lamella. The structural and chemical properties of the laser epilayer were analyzed using high-angle annular dark-field scanning transmission electron microscopy (HAADF-STEM) and high-resolution transmission electron microscopy (HRTEM) with a Talos F200s microscope operating at 200 kV. EDS elemental maps were obtained using a dual-quadrant Super-X EDS detector.

**Device fabrication**

The InAs/GaAs quantum dot epitaxy wafer was grown and then processed into broad-area Fabry-Perot (FP) lasers with widths of 50 μm. This was done using standard photolithography and wet etching techniques with phosphoric acid etchant (phosphoric acid: hydrogen peroxide: water = 1: 1: 6) down to the middle of the upper waveguide layer. A 350-nm-thick $SiO_2$ passivation layer was deposited on the sample surface using plasma-enhanced chemical vapor deposition (PECVD). Electrode windows were patterned using photolithography and buffered oxide etch (BOE) solution on the ridge top. Next, p-type electrode stripes were fabricated by depositing Ti/Au onto the top of the mesa using electron beam evaporation, followed by a lift-off process. The substrate was thinned to 150 μm through chemical mechanical planarization. Then, n-type electrodes (Ge/Au/Ni/Au) were deposited on the backside of the substrate. A rapid post-annealing process was carried out at 380 °C for 45 seconds to achieve ohmic contact between the metal and the semiconductor. The processed wafer was cleaved without facet coating into laser bars. Finally, the bars were mounted on indium-plated copper heat sinks with the epitaxial side up and connected to copper strips using gold wire bonding for testing.

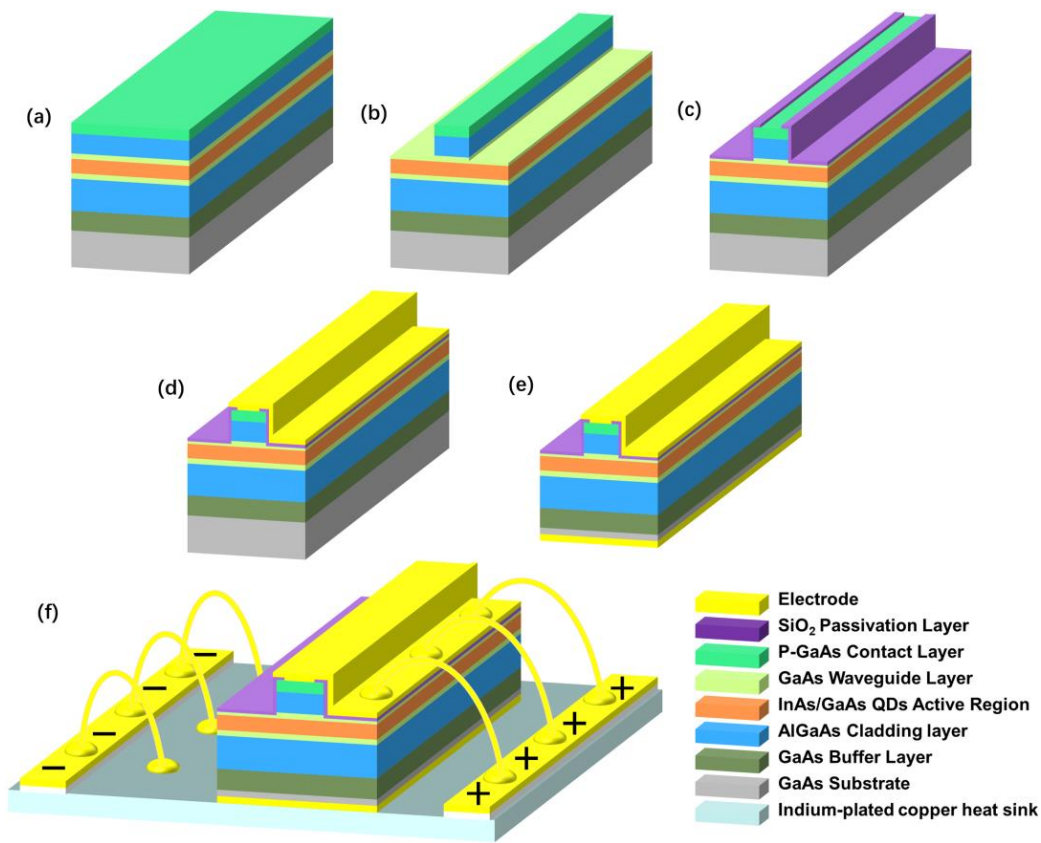

**Figure 6**. (a) laser structure epitaxy; (b) patterning using photolithography, followed by wet etching to define ridges; (c) deposition of $SiO_2$ passivation layer, followed by wet etching to open electrode window; (d) p-contact metallization (Ti/Au); (e) substrate thinning and n-contact metallization (Ge/Au/Ni/Au) followed by annealing; (f) wafer cleave, bars mounting and gold wire bonding.

**Measurement setup**

The QD laser devices were measured at 25°C using a continuous wave mode electrical drive with a source meter (Keithley, 2450). The optical output power was measured using an optical power meter (Newport, 1938-R) with an integrating sphere detector (Newport, 819D-IG-2-CAL2). The electroluminescence spectra of the LD were measured using an optical spectrum analyzer (Yokogawa, AQ6370D). Additionally, the near-field intensity profile was measured using a slit beam profiler (Thorlabs BP209-IR2/M).

**Hardware wiring scheme**

Our model runs on a Windows 10 operating system with an AMD R9 7950X CPU, 64 GB of RAM, and an NVIDIA 3090 graphics card. The 2 TB SSD is split into a system and data storage drive to allow the program to carry out tasks and store real-time collected RHEED data using different paths. This helps prevent overheating during high-load operation, reducing the risk of malfunctions. The program automatically collects information from the temperature controller, shutter controller, and camera. The temperature controller is connected via the Modbus protocol and utilizes an address lookup function to deploy commands during program execution. The shutter controller manages the states of shutters through a multi-bit binary string. The camera connects via a USB 3.0 interface and is pre-configured using a third-party camera driver with parameters like exposure time, color balance, and brightness. This setup compensates for the fluorescent screen's green colour, ensuring effective signal capture by the camera. This integrated system enables efficient real-time monitoring and analysis of the growth process.

**Model construction environment**

The model was developed in the Jupyter Notebook environment on an Ubuntu system using Python version 3.9 to ensure compatibility with various libraries and tools essential for ML. We installed PyTorch, optimized explicitly for CUDA 11.8, to leverage the computational power of our NVIDIA GPU for accelerated training and inference programs. To enhance model robustness, we implemented data preprocessing scripts that quickly converted the original video data into NumPy arrays. These scripts also incorporated random data augmentation techniques such as rotation, flipping, and colour adjustments within the Ubuntu environment. This augmentation enriched the dataset and improved generalization during model training. The processed data was subsequently organized and stored on our computer's SSD, with appropriate directory structures established for efficient access and retrieval during the training phase.

**Program interface and deployment environment**

The program uses LabVIEW and integrates NI VISA, NI VISION, and Python libraries for data acquisition and processing. It employs ONNX for model deployment and TensorRT for accelerated inference. Initially, the software adjusts the temperature and shutter states to grow a thickness of GaAs. Then, the temperatures for InAs and LT-GaAs growth are adjusted. During the InAs growth, the main shutter closes, entering an overshoot phase to ensure a uniform substrate surface before starting the growth of InAs QDs. The program then switches the shutter state to begin InAs QD growth while analyzing real-time RHEED data. Model outputs are displayed at the top of the interface as labels, and the material growth status is indicated in a "Reminder information". After completing the growth of each layer, the system automatically transitions to the next layer, requiring only manual adjustments to the RHEED shutter state and the As needle valve throughout the process.

Data Availability

The datasets generated during and/or analysed during the current study are available in the Figshare repository. Source data are provided with this paper.

Code Availability

The codes supporting the findings of this study are available from the corresponding authors upon request.

## Acknowledgements

This work was supported by the National Key R&D Program of China (Grant No. 2023YFB2805900, S. M. C.; 2021YFB2206500, C. Z.), National Natural Science Foundation of China (Grant No. 62274159, C. Z.), the "Strategic Priority Research Program" of the Chinese Academy of Sciences (Grant No. XDB43010102, C. Z.), and CAS Project for Young Scientists in Basic Research (Grant No. YSBR-056, C. Z.).

## Author Contributions Statement

C. S. and W. K. Z. contributed equally. C. Z. conceived the idea, designed the investigations and the growth experiments. C. S. and W. K. Z. performed the molecular beam epitaxial growth. W. K. Z. and H. Y. H fabricated the laser. C. S., K. Y. X., S. J. P, H. C., C. X., and S. M. C did the sample characterization and device test. C. S., T. K. N, C. L. X., F. Q. L, B. S. O, and C. Z. wrote the manuscript. C. Z. led the molecular beam epitaxy program. B. X. and Z. G. W. supervised the team. All authors have read, contributed to, and approved the final version of the manuscript.

## Competing Interests Statement

The authors declare no competing interests.

## Supplementary Information

Details on the results of InAs QDs growth under different V/III ratios, the result of InAs QDs growth under different deposition amounts, the detailed information on modules in the GLAM block, the training and validation speed of the two models, the experiment setup and program interface, the setting of the substrate temperature ramp rate, the results of control sample A, the results of control sample B, the data of single-layer buried QDs , the results of control sample C , the TEM images of 5-layer InAs QDs, the two-dimensional and three-dimensional near-field intensity profile. The video demonstrated the sample of 5-layer buried QDs for the laser during growth.